\documentclass[twocolumn,tighten,twocolappendix]{aastex631}
\usepackage{graphicx}
\usepackage{url}
\usepackage{color}
\usepackage{multirow}
\usepackage{ragged2e}
\usepackage{hyperref}
\usepackage{url}
\usepackage{amsmath} 
\usepackage{booktabs}
\usepackage{comment}
\usepackage{afterpage}
\usepackage{mathtools}
\usepackage{tabularx}
\usepackage{bold-extra}
\usepackage{xspace}
\usepackage{relsize}
\usepackage{newunicodechar}
\usepackage{lipsum}
\usepackage{eht}

\usepackage[encapsulated]{CJK}
\usepackage{ucs}
\usepackage[utf8x]{inputenc}

\usepackage{eht}

\def\VirgoAstar{M87*\xspace}
\def\SgrAstar{Sgr A*\xspace}
\def\emcee{\texttt{emcee}\xspace}

\begin{document}

\title{Constraining the Existence of Axion Clouds in \VirgoAstar with Closure Trace Analyses}

\shorttitle{Constraining the Existence of Axion Clouds in \VirgoAstar with Closure Trace Analyses}

\author[0009-0004-9417-2213]{Zhiren Wang}
\affiliation{Perimeter Institute for Theoretical Physics, 31 Caroline Street North, Waterloo, ON, N2L 2Y5, Canada}
\affiliation{Department of Physics and Astronomy, University of Waterloo, 200 University Avenue West, Waterloo, ON, N2L 3G1, Canada}
\affiliation{Waterloo Centre for Astrophysics, University of Waterloo, Waterloo, ON, N2L 3G1, Canada}

\author[0000-0002-3351-760X]{Avery E. Broderick}
\affiliation{Perimeter Institute for Theoretical Physics, 31 Caroline Street North, Waterloo, ON, N2L 2Y5, Canada}
\affiliation{Department of Physics and Astronomy, University of Waterloo, 200 University Avenue West, Waterloo, ON, N2L 3G1, Canada}
\affiliation{Waterloo Centre for Astrophysics, University of Waterloo, Waterloo, ON, N2L 3G1, Canada}

\begin{abstract}
    Black holes can amplify incoming bosonic waves via rotational superradiance, inducing bound states of ultralight bosons around them. This phenomenon has the potential to confine the parameter spaces of new bosons. Axions and axion-like particles (ALPs) are candidate beyond-standard-model particles that can form such clouds around supermassive black holes (SMBHs) and impact the polarization signal in a similar fashion to Faraday rotation via axion-photon coupling. Prior efforts have used polarized images from the Event Horizon Telescope (EHT) M87 2017 observations to limit the dimensionless axion-photon coupling to previously unexplored regions. However, with the novel calibration-insensitive quantities, closure traces and conjugate closure trace products, it is possible to constrain the existence of axion clouds while avoiding the dominant sources of systematic uncertainties, e.g., station gains and polarization leakages. We utilize a simple geometric model for the polarization map of M87$^*$ to fit the model parameters with both simulated and real data sets and reach a comparable level of constraint in the accuracy with which an axion cloud may be excluded in M87. Future applications of our approach include subsequent \VirgoAstar and \SgrAstar observations by EHT and next-generation EHT (ngEHT) are expected to produce stronger constraints across a wider range of axion and ALP masses. Because it does not require imaging, closure trace analyses may be applied to target AGN for which imaging is marginal, extending the number of SMBHs from which axion limits may be obtained significantly. 
\end{abstract}



\nocite{M87PaperI}
\nocite{M87PaperII}
\nocite{M87PaperIII}
\nocite{M87PaperIV}
\nocite{M87PaperV}
\nocite{M87PaperVI}
\nocite{M87PaperVII}
\nocite{M87PaperVIII}

\section{Introduction} 
\label{sec:Intro}

\begin{figure*}
    \includegraphics[width=\textwidth]{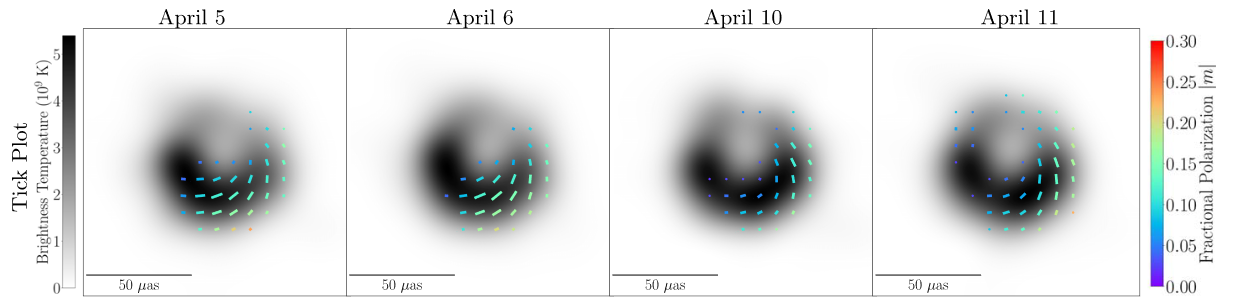}
    \caption{Time evolution of the fiducial average polarimetric images of \VirgoAstar from EHT 2017 Campaign from Apr 5 to Apr 11. Total intensity is shown in grayscale. EVPA is shown in colorful ticks, while the length indicates linear polarization intensity magnitude, and color indicates fractional linear polarization. (Reproduced from \citetalias{M87PaperVII}.) }
    \label{fig:VII}
\end{figure*}

\begin{figure}
    \includegraphics[width=\columnwidth]{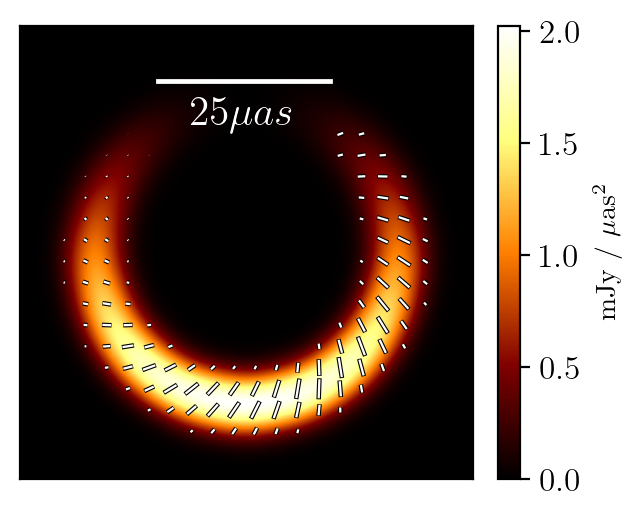}
    \caption{Stationary polarimetric model image before the inclusion of an axion cloud (i.e., $\theta_1=0$).  Length of the polarization ticks indicate the polarized flux, $|\mathcal{P}|$.  The location of the Stokes $\mathcal{I}$ and $|\mathcal{P}|$ maxima, and $\theta_0$ are chosen to approximate the morphologies seen in \autoref{fig:VII}.}
    \label{fig:Radial}
\end{figure}

\begin{figure*}
    \includegraphics[width=\textwidth]{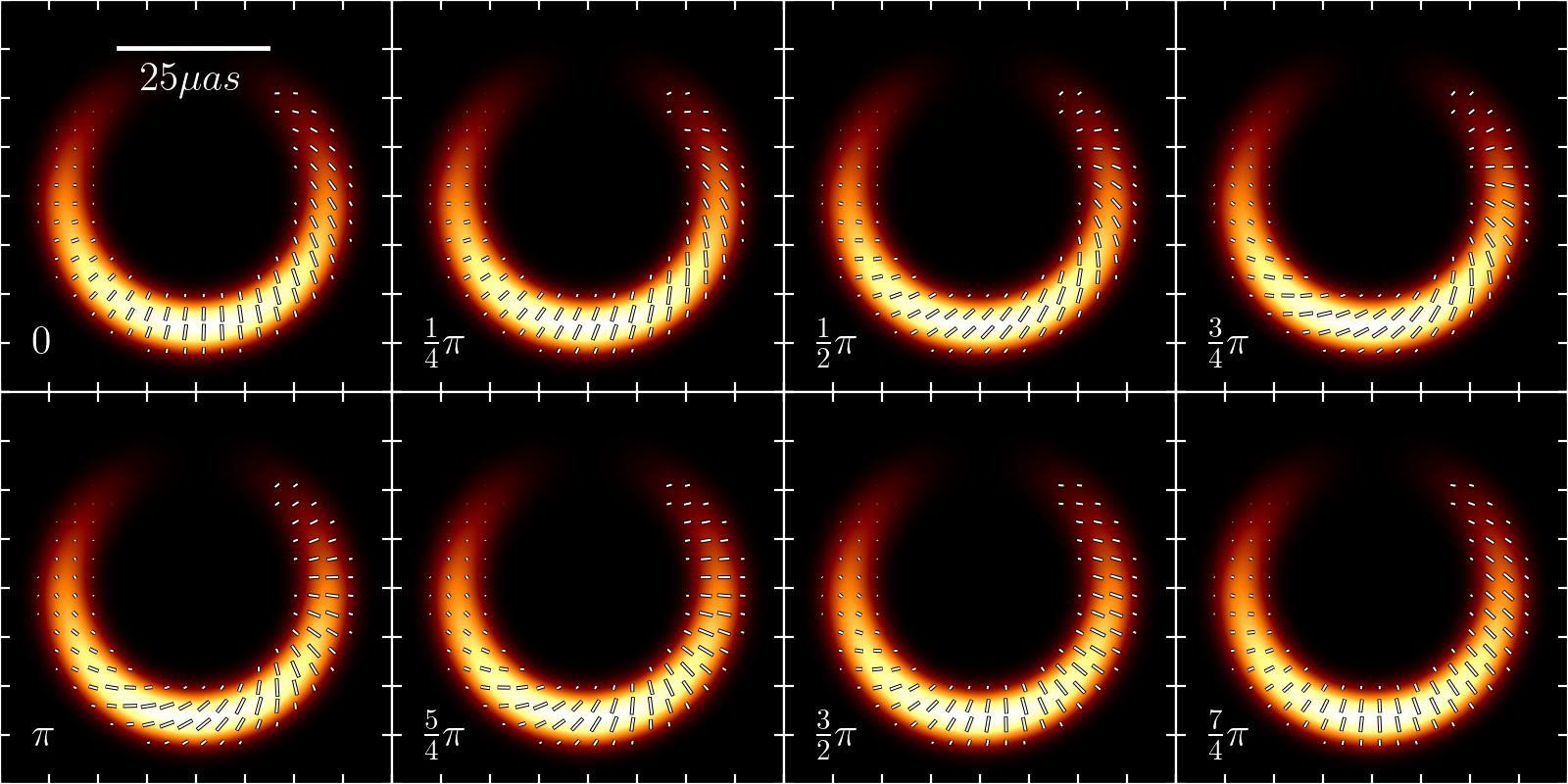}
    \caption{Eight snapshots with phases between 0 and 2$\pi$ captured 
    from the time-dependent model, approximating the stationary model in \autoref{fig:Radial} with a time-dependent axion cloud. The phase of each snapshot is labeled at the bottom left. The position angles of maximal Stokes $\mathcal{I}$ and polarization flux are chosen to match the actual image of 2017 Apr 11.  The intensity color range and the length of the polarization ticks are the same as in \autoref{fig:Radial}}
    \label{fig:eight frame}
\end{figure*}

Being among the most luminous astronomical objects in the universe, active galactic nuclei (AGN) are observed and studied across the entire electromagnetic spectrum. For example, in the radio regime, AGN are identified as extragalactic sources of synchrotron emission, which arises from the interaction of relativistic electrons with magnetic fields. The radio emission from AGN is often highly variable and can be used to study the properties of the relativistic outflows that are launched from the vicinity of the SMBH \citep{Boccardi2017}. Furthermore, new technologies in radio observations of AGNs are constantly pushing the limit of the resolution of images of those cosmic giants, revealing their complicated magnetic structures. 

Among the multiple pathways in the radio study of AGN, polarization observation of those targets and their central engines plays a crucial role in unveiling the inner structures, especially the intrinsic magnetic fields of these cosmic giants [\citealp{Johnson2015}, \citetalias{M87PaperVII}]. Fortunately, the advent of very long baseline interferometry (VLBI) has enabled us to overcome the challenge posed by their extremely small angular sizes, allowing for detailed analyses of their vicinity \citep{interferometry3rdversion}. Polarization signals from those massive giants encode essential information about the dynamics and magnetic structures in the surrounding environments. In 2021, the Event Horizon Telescope Collaboration presented the first horizon-resolving polarimetric images of \VirgoAstar shadow \citepalias{M87PaperVII}, enabling unprecedented detailed analysis of its physics. 

Besides the study of magnetic structures \citepalias{M87PaperVIII}, the acceleration of the bulk flow, and the relativistic heating of jet constituents \citep{acceleration}, observations of SMBHs offer a unique opportunity to probe beyond-standard model (BSM) physics and place constraints on speculative particles, such as QCD (quantum chromodynamics) axions, which have been proposed as a solution to the strong CP (charge-parity) problem \citep{Peccei&Quinn1977}, and axion like particles (ALPs), which serve as candidates for dark matter \citep{Preskill1983}. 

The underlying mechanism is simple as follows. Axions would accumulate around black holes via black hole superradiance, under which traveling light bosons form bound states around SMBHs and grow exponentially as long as superradiance condition, $\omega < m\Omega_H$, is satisfied. $\omega$ is the frequency of the incoming wave, $\Omega_H$ is the angular frequency of the black hole and $m$ is the azimuthal quantum number. In addition, the most significant superradiant rate occurs when $m=1$, the lowest possible energy state that satisfies the condition \citep{Azimuthal}. The superradiant rate is maximized when the axion's reduced Compton wavelength $\lambda_c$ is comparable to the black hole size \citep{Arvanitaki2015}. In the presence of a background axion field, the modification of a photon's equation of motion due to the axion-photon interaction will introduce a periodic oscillation to the EVPAs (electric vector position angles) of linearly polarized photons \citep{Carroll1990,Carroll1991,Harari1992}. Theoretically, modelling the variations of EVPAs around the black hole (BH) shadow of an SMBH should indicate novel constraints on light axions aside from previous results\footnote{For an extensive compilation of existing experimental axion constraints from the literature, ecompassing a broad range of potential masses, see \href{https://cajohare.github.io/AxionLimits/docs/ap.html}{cajohare.github.io/AxionLimits/docs/ap.html} \citep{AxionLimits}.}. 

An experiment was then proposed when \citet{Chen2020} addressed that this axion-induced birefringence effect can cause the EVPAs on the photon ring of an SMBH to vary with time and position. In the condition of photon propagating in a background axion field, axion-photon interaction modifies the Lagrangian as
\begin{equation}
    \mathcal{L} = -\frac{1}{4}F_{\mu\nu}F^{\mu\nu} - \frac{1}{2}g_{a\gamma}aF_{\mu\nu}\Tilde{F}^{\mu\nu} + \frac{1}{2}\nabla^{\mu}a\nabla_{\mu}a - V(a),
\end{equation}
in which $g_{a\gamma}$ is the axion-photon coupling constant, $F_{\mu\nu}$ is the electromagnetic tensor and $\Tilde{F}^{\mu\nu}$ is its dual tensor. A naive deduction would be that this axion-induced effect would function very similarly to the normal Faraday effect. Successive derivations with the assumption that the photon frequency is much larger than axion's reduced Compton wavelength $\lambda_c$ reached a simple expression for the EVPA rotation of a linearly polarized photon:
\begin{equation}
    \Delta \theta = g_{a\gamma}[a(t_{\rm obs},\boldsymbol{x}_{\rm obs})-a(t_{\rm emit},\boldsymbol{x}_{\rm emit})]
\end{equation}
The rotation only depends on the difference between the emitting and observing axion field values. With further appropriate assumptions and simplifications that describe the EVPA variation in Boyer-Lindquist coordinates of the black hole, the final form can be parametrized like
\begin{equation}
    \Delta \theta(t,\phi) = \mathcal{A}(\phi)\cos[{\omega t + \phi + \beta(\phi})],
\end{equation}
in which a polar coordinate ($r,\phi$) centered at the black hole shadow is used. For a detailed derivation of the previous equations, see \citet{Chen2022}.

Recently, a practical attempt by the same group \citep{Chen2022a} utilize the polarimetric information of \VirgoAstar \citepalias{M87PaperVII} to reach a previously unexplored parameter space of light axions. Specifically, they implement constraints on the joint distribution of the axion mass $m_a$ and the dimensionless axion-photon coupling constant $c_{a\gamma} \equiv 2\pi g_{a\gamma}f_a$. However, one caveat is that the parametrized form of EVPA variation relies on the assumption of a saturated axion cloud. While it is impossible to know if the axion cloud is saturated given the unknown recent astrophysical history of M87, we adopt this assumption henceforth (see \citet{Chen2022} for a full discussion of this issue).

Despite its many advantages, polarized radio interferometry, which is used to generate the polarimetric images of \VirgoAstar, suffers from multiple systematic uncertainties, including station gains that modulate both the amplitude and the phase of the observed signal, and polarimetric leakages from one channel to another, e.g., from right-hand circular polariztion to left-hand circular polarization \citepalias{M87PaperIII}. The former one can be circumvented by "closure" quantities (closure phase and closure amplitude) on multiple sites that are immune to station-based effect because station-based gain errors are cancelled out \citep{CP,CA,Readhead}, while the latter one is calibrated with suboptimal assumptions \citepalias{M87PaperVII}. Luckily, \citet{BroderickPesce_2020} introduced a novel calibration-insensitive closure quantity: closure trace (CT), defined on station quadrangles, that is insensitive to station gains and polarimetric leakages (D-terms) and comprises all calibration-independent information in the visibility data; and a subsequent quantity, the conjugate closure trace product (CCTP), that can serve as a robust indicator of polarization sources. 

Therefore, based on the information of axion-photon coupling and CT analysis, we present a new method to constrain light axions and demonstrate that it can reach a roughly similar level of constraint as \citet{Chen2022a}. Our method, being insensitive to a much larger class of errors than ameliorated by gains and D-terms, circumvents the time-consuming direct calibration process to the final polarized images of \VirgoAstar and is complete in retrieving polarimetric information. This novel method can be further applied to future EHT and next-generation EHT (ngEHT) observations \citep{ngEHT}, for which more quadrangles will be available, to reach stronger constraints on axions, ALPs and other ultralight bosons. 

The structure of this paper is as follows. In \autoref{sec:Geo Model}, we introduce our model of EVPA variations, explore and discuss its feasibility. In \autoref{sec:MCMC}, we present our Monte Carlo Markov Chain (MCMC) fits for the model and its implications. Finally, in \autoref{sec:Conclusion} we make conclusions.

\section{A Simple Geometric Model for the Linear Polarization Maps of \VirgoAstar}
\label{sec:Geo Model}

\begin{figure}
    \includegraphics[width=\columnwidth]{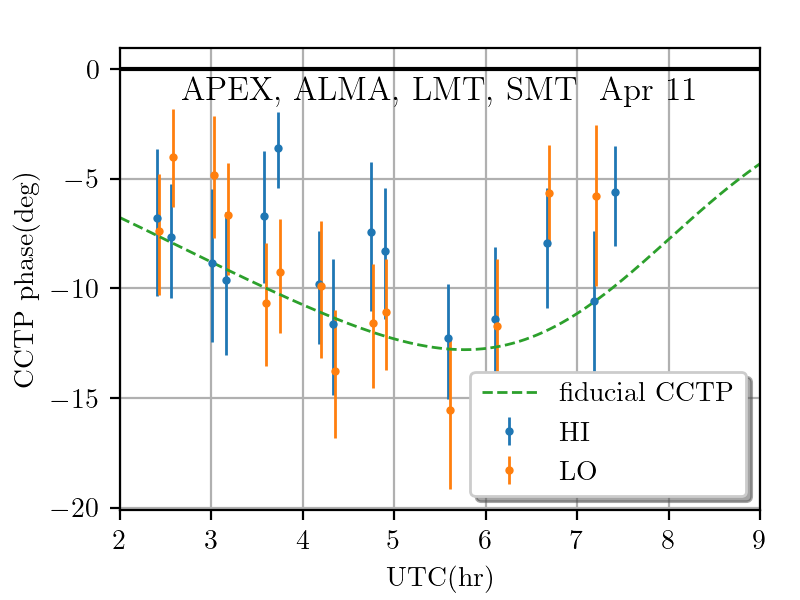}
    \caption{HI+LO band real CCTP measurements for 2017 Apr 11, with CCTP curves from the fiducial set of parameters. HI and LO refer to the high frequency band and low frequency band centered at 229.1 GHz and 227.1 GHz, respectively. LO band data are intentionally shifted to the right by 0.02 hrs for better readability. The errorbars in the CCTP measurements are estimated using Monte Carlo sampling, are purely due to the underlying thermal errors in the visibilities, and are well-approximated as Gaussian errors. }
    \label{fig:Fiducial}
\end{figure}

\begin{figure}
    \includegraphics[width=\columnwidth]{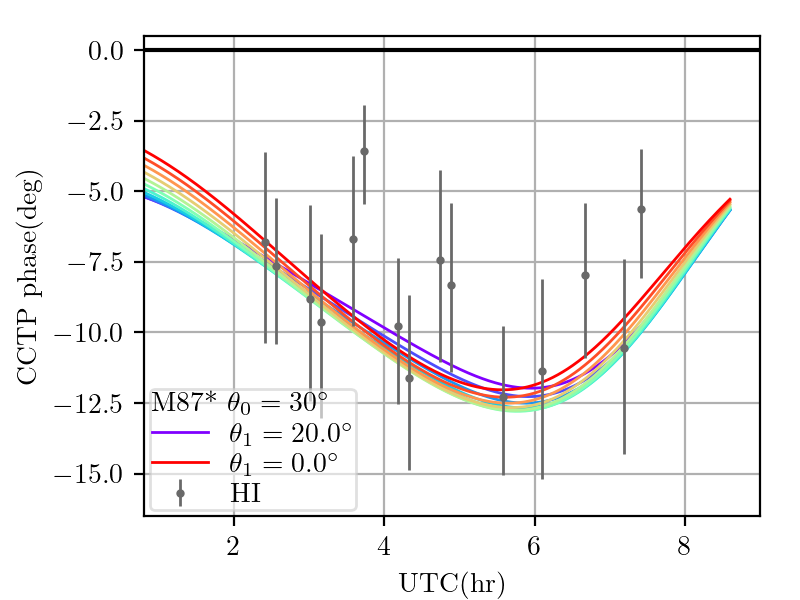}
    \caption{Change in $\theta_1$, the variable EVPA pitch angle proportional to the axion-photon coupling constant. The first line in the legend indicates the target of the simulated CCTP. $\theta_0=30^{\circ}$ is arbitrary as changes of $\theta_0$, the mean EVPA pitch angle, have no impact on the shape of the curve. From red to purple, the spacing of $\theta_1$ between two consecutive colors is $2^{\circ}$. For reference, the HI band real CCTP measurements on Apr 11 are plotted in light grey.}
    \label{fig:a0}
\end{figure}

The most striking feature of the EHT images of \VirgoAstar (\autoref{fig:VII}) is the clear presence of a ring, presumably encircling the black hole shadow. Upon this ring lies a ``twisty'' linear polarization pattern with the EVPA exhibiting a nearly linear dependence on azimuth. Given these gross features, we generate a simple geometric model for the polarized images of \VirgoAstar with which to explore the constraints made possible using the measured CCTPs.

\subsection{Image Domain Model}

Besides the outstanding feature of an annulus in the EHT images of \VirgoAstar, we clearly notice the position angle of the maximum Stokes $\mathcal{I}$ map lies in the southern direction, the position angle of the maximum linear polarization lies in the SW direction and the EVPA position angle around the shadow appears to deviate from an overall radial EVPA pattern (\autoref{fig:VII}). Based on this information and putting time-dependent EVPA variations aside, we begin by creating movie objects to model the flux and the polarization map, that can be processed by \ehtim \citep{ehtim_paper}. The parameters include the diameter $d$ of the annular emission region in $\mu as$, annulus width $w$, linear polarization fraction $m$, the number of periods $n_{\rm cyc}$ per 24 {\rm h}, the position angle $\Phi_{\rm I}$ of the maximum brightness in Stokes $\mathcal{I}$ map and the position angle $\Phi_{\rm pol}$ of the maximum polarized flux, where the latter two angles start from 0 in the south direction on the image and increase clockwise, i.e., 
\begin{equation}
\begin{multlined}
    \mathcal{I}(r,\phi;\mathcal{I}_0,d,w,\Phi_{\rm I}) =\\
    \mathcal{I}_0 \begin{cases}
         \cos^2\left[(\phi+\Phi_{\rm I})/2\right] & \text{if}~d-w < r < d \\
         0 & \text{otherwise,}
    \end{cases}
\end{multlined}
\end{equation}
and
\begin{equation}
\begin{multlined}
    |\mathcal{P}|(r,\phi;\mathcal{I}_0,d,w,\Phi_{\rm pol}) = \\
    m \mathcal{I}_0 \begin{cases}
         \cos^2\left[(\phi+\Phi_{\rm pol})/2\right] & \text{if}~d-w < r < d \\
         0 & \text{otherwise.}
    \end{cases}
\end{multlined}
\end{equation}
Given an EVPA, $\theta(t,\phi)$, we can then construct the remaining Stokes parameters via
\begin{equation}
\begin{aligned}
    &\mathcal{Q}(r,\phi;\mathcal{I}_0,d,w,\Phi_{\rm pol},\theta) \\
    &\qquad = |\mathcal{P}|(r,\phi;\mathcal{I}_0,d,w,\Phi_{\rm pol}) \cos\left[2\theta(t,\phi)\right]\\
    &\mathcal{U}(r,\phi;\mathcal{I}_0,d,w,\Phi_{\rm pol},\theta) \\
    &\qquad = |\mathcal{P}|(r,\phi;\mathcal{I}_0,d,w,\Phi_{\rm pol}) \sin\left[2\theta(t,\phi)\right].
\end{aligned}
\end{equation}

We convolve the polarimetric images by a Gaussian filter and then generate movies that reflect the time evolution of EVPAs that results from axion-photon coupling. Combining the most prominent features of the polarized images of \VirgoAstar and the parametrization of EVPA variation by \citet{Chen2022a}, we construct a simple geometric model of the EVPA oscillation by introducing an additional time-dependent position angle $\theta$ on top of a fundamental time-independent radial EVPA pattern (\autoref{fig:Radial}) around the shadow. 
\begin{equation}
    \theta(t,\phi) = \theta_0 + \theta_1\cos{[\omega(t - r_{\rm ring}\sin{17^{\circ}}\cos{\phi}) + \phi + \delta]}
    \label{eq:theta}
\end{equation}
In this model, $\theta_1$ is the oscillation amplitude and is proportional to $c_{a\gamma}$ and $\omega$ is the axion angular frequency proportional to $m_a$, later substituted by $n_{cyc}$, as in \autoref{tab:params}. The rest of the parameters are defined as follows: $r_{\rm ring} = \sqrt{27}r_{\rm g}$ is the radius of the shadow (assuming a=0) of \VirgoAstar, $\theta_0$ is the initial value of $\theta$, $\phi$ is the azimuthal angle on the annulus and $\delta$ describes an arbitrary initial phase. The inclination angle of $17^{\circ}$ between the spin direction of \VirgoAstar and the sky plane is introduced by \citet{Chen2022a} to account for the time delay of emissions. $\theta_1$ is independent of the azimuthal angle because the azimuthal variation is rather marginal, as seen in \citet{Chen2022a}. \autoref{fig:Radial} shown below is the fundamental radial EVPA plot we started from, while in \autoref{fig:eight frame} we added $\theta$ and clipped a total of eight frames from the EVPA movie of our model with morphologically estimated parameters of 2017 April 11 to fully cover its evolution. This simple geometric model captures the characteristics of time-dependent EVPA variation induced by axion-photon coupling appearing as propagating waves around the shadow. 

\citet{Chen2022a} extended the "forbidden" region of axions on the $c_{a\gamma}$-$m_a$ parameter space by directly fitting the differential EVPA on the polarimetric images across 2017 April 5, 6, 10, 11, highlighting the possibility of exploring new physics with the unprecedented EHT images on SMBHs (although assuming the existence of a saturated, superradiantly generated axion cloud around \VirgoAstar). However, it is important to note that this approach neglect the systematic uncertainties within the EVPA, like station gains and D-terms, which will be discussed below, associated with calibrations to the final polarimetric images. Our new method, on the other hand, takes advantage of novel calibration-insensitive quantities for radio astronomy, to "observe" the time-dependent movies generated by this model, and reach an approximately comparable level of constraints on light axions, without the need for full polarimetric image reconstruction, with the attendant systematic uncertainties. 

\begin{figure}
    \includegraphics[width=\columnwidth]{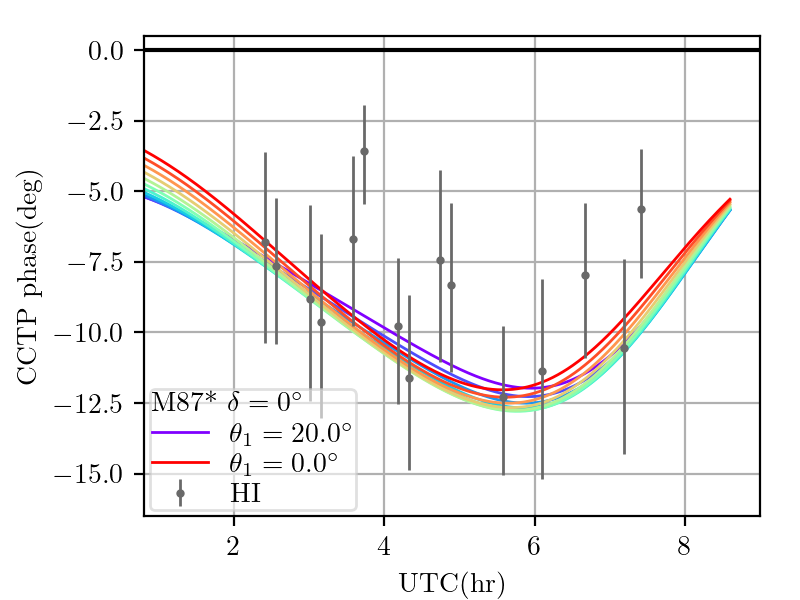}
    \includegraphics[width=\columnwidth]{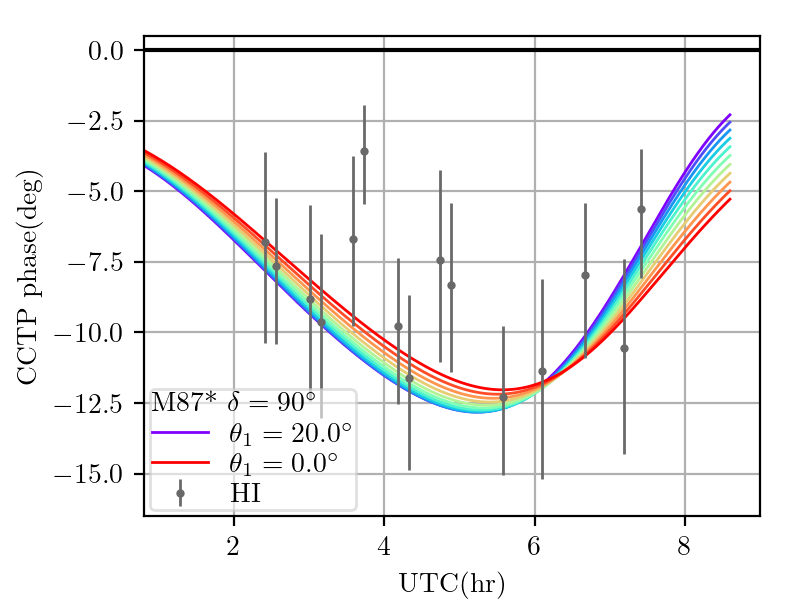}
    \includegraphics[width=\columnwidth]{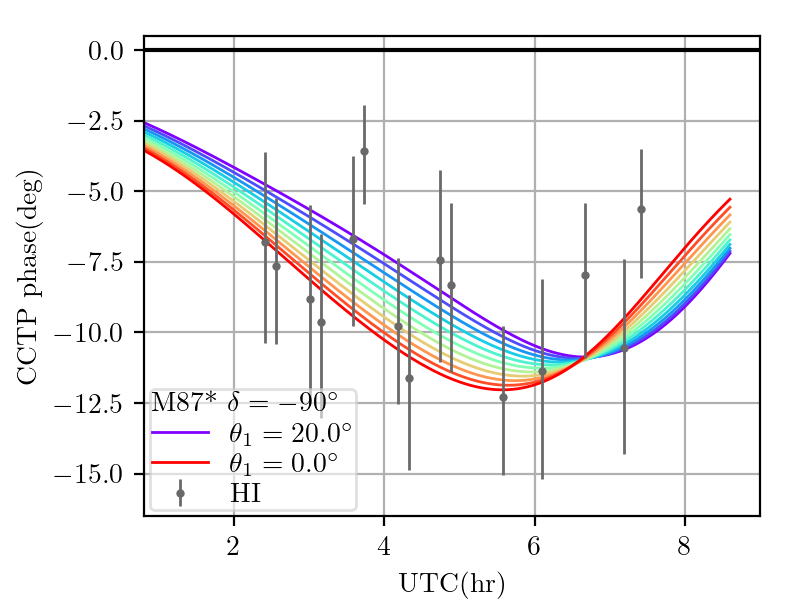}
    \caption{Change in $\delta$, the axion cloud phase offset. The first line in the legend indicates the target of the simulated CCTP and the parameter to be varied. From red to purple, the spacing of $\theta_1$ between two consecutive colors is $2^{\circ}$. For reference, the HI band real CCTP measurements on Apr 11 are plotted in light grey.}
    \label{fig:delta}
\end{figure}

\begin{figure*}
    \includegraphics[width=0.5\textwidth]{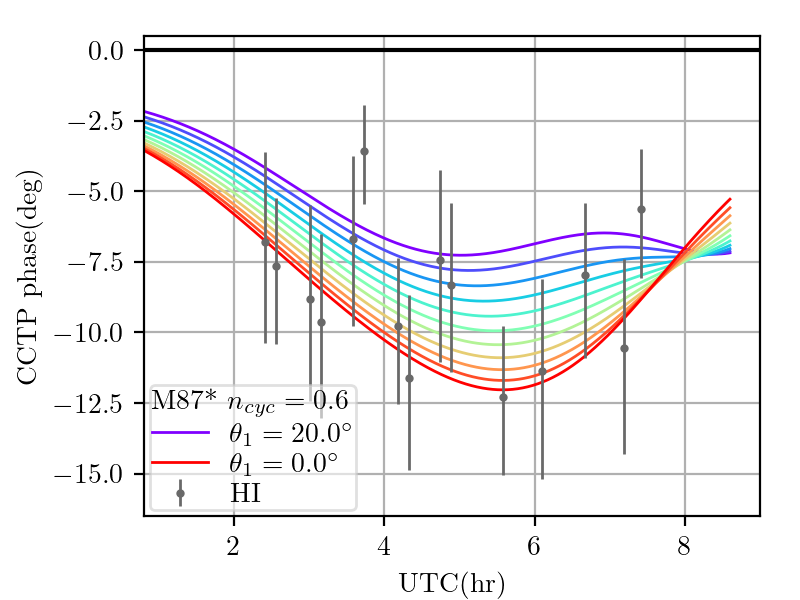}
    \includegraphics[width=0.5\textwidth]{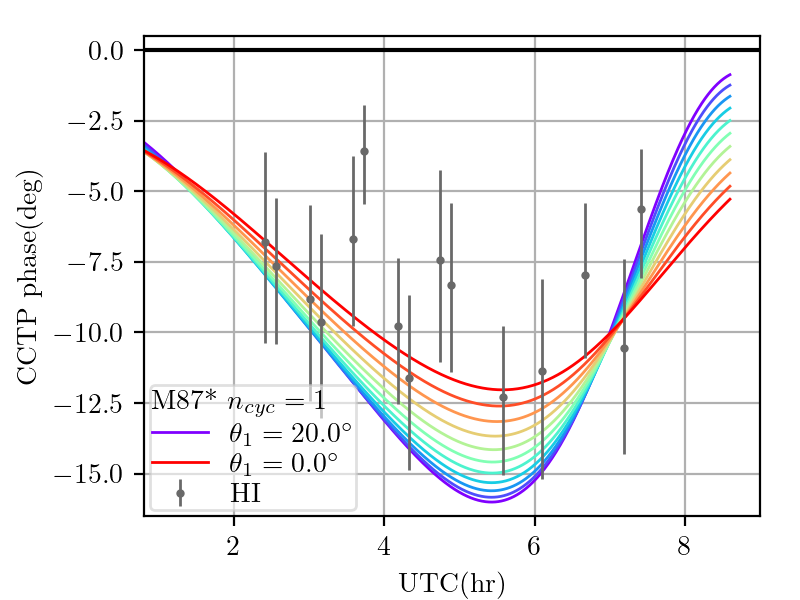}
    \includegraphics[width=0.5\textwidth]{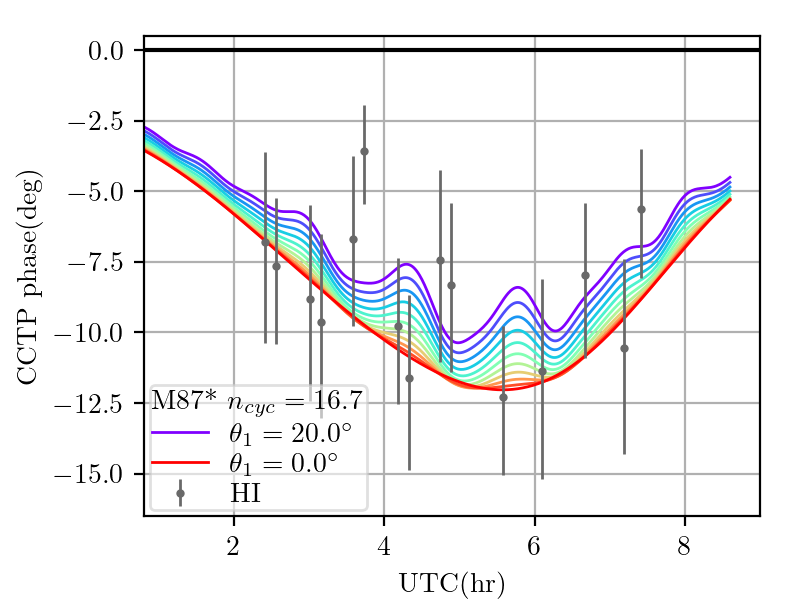}
    \includegraphics[width=0.5\textwidth]{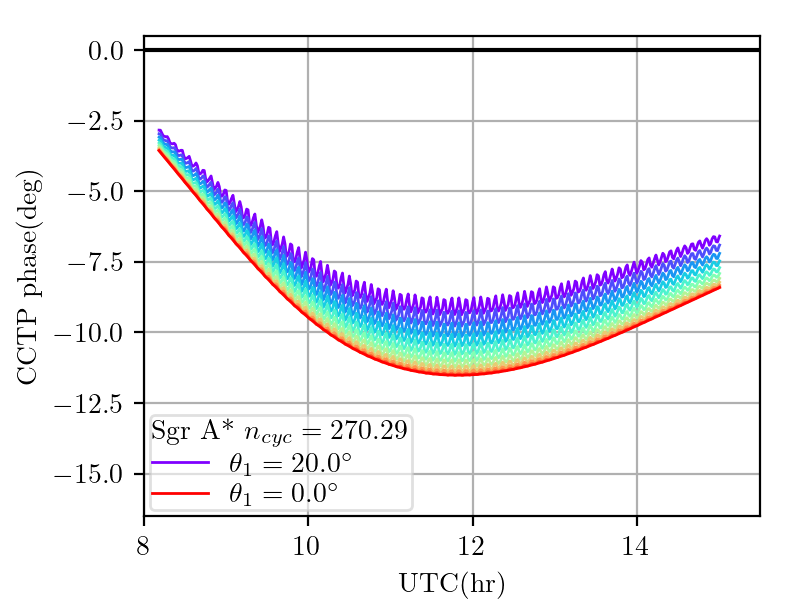}
    \caption{Change in $n_{\rm cyc}$, or the number of cycles which is proportional to axion mass. The first line in the legend indicates the target of the simulated CCTP and the parameter to be varied. From red to purple, the spacing of $\theta_1$ between two consecutive colors is $2^{\circ}$. For reference, the HI band real CCTP measurements on Apr 11 are plotted in light grey.}
    \label{fig:ncycles}
\end{figure*}

\subsection{Constructing EHT Observables}

The primary EHT data products are visibilities, corresponding to the Fourier transform of the Stokes $\mathcal{I}$, $\mathcal{Q}$, $\mathcal{U}$, and $\mathcal{V}$ brightness maps.  We obtain these using the \ehtim package via the $observe$ function, generating visibilities in the more common circular basis, $RR$, $LL$, $RL$, $LR$ \citepalias[see][]{M87PaperVII}, commonly collected into a coherence matrix,
\begin{equation}
    \boldsymbol{V}_{ij}=
    \left(\begin{array}{cc}
        RR_{ij} & RL_{ij} \\
        LR_{ij} & LL_{ij} \\
    \end{array}\right),
\end{equation}
where the indicies $ij$ refer to the two stations defining a particular baseline.  All baselines relevant for EHT are generated at the same time. However, the observed $\boldsymbol{V}_{ij}$ are degraded by aforementioned station gains and polarimetric leakages (D-terms), which means the true coherence matrix are modified as
\begin{equation}
    \boldsymbol{V}_{ij} = \boldsymbol{G}_A\boldsymbol{D}_A \boldsymbol{\bar{V}}_{ij}\boldsymbol{D}^{\dagger}_B\boldsymbol{G}^{\dagger}_B,
\end{equation}
where $\boldsymbol{G}$ and $\boldsymbol{D}$ contain the gain terms and leakage terms for each station respectively, and $\dagger$ is the complex conjugate transpose. Polarimetric imaging requires full reconstructions of the time-dependent station characteristics (i.e.,  the $\boldsymbol{G}$ and $\boldsymbol{D}$).  However, the immunity/insensitivity of the CT and CCTP to $\boldsymbol{G}$ and $\boldsymbol{D}$ present an opportunity to avoid this process.

The CTs are generated from combinations of $\boldsymbol{V}_{ij}$ on quadrangles, sets of four stations whose baselines form a closed circuit \citep{BroderickPesce_2020}. 
\begin{equation}
    \mathcal{T}_{ijkl}
    =
    \frac{1}{2} {\rm Tr}\left(
    \boldsymbol{V}_{ij} \boldsymbol{V}^{-1}_{kj}
    \boldsymbol{V}_{kl}
    \boldsymbol{V}^{-1}_{li}
    \right).
\end{equation}
These are invariant to all linear, station-based corruptions of the coherence matrices, including atmospheric phase delays, receiver gains, and polarization leakage.  The full complement of $\mathcal{T}_{ijkl}$ contain all of the residual information in the coherence matrices apart from the standard calibration quantities, and are a superset of the more familiar closure phases and closure amplitudes \citep{interferometry3rdversion}.  

While the $\mathcal{T}_{ijkl}$ are generally sensitive to polarized and unpolarized image substructure, a specific combination of them, the CCTP defined by
\begin{equation}
    \mathcal{C}_{ijkl} = \mathcal{T}_{ijkl} \mathcal{T}_{ilkj},
\end{equation}
are sensitive only to polarization substructure; in the absence of polarization substructure, the CCTP is identically unity.  Therefore, the CCTPs are valuable probes of polarization in much the same way that closure phases are sensitive to asymmetry.   

For \VirgoAstar, the only quadrangle exhibiting CCTPs with the most significant departures from unity, and therefore the most significant evidence for polarized structure, was that combining APEX, ALMA, LMT, and SMT \citepalias[see Fig 13 of][]{M87PaperVII}.  One potential explanation is that ALMA and LMT have very larger apertures, and are thus the most sensitive stations. Another possible reason is that the ordered polarization pattern indicates that the most power is on scales of a few G$\lambda$, which corresponds to the Chile(ALMA, APEX)-LMT and Chile-SMT baselines (usually below 4 G$\lambda$). Therefore, we construct the CCTPs from the complex visibilities generated by \ehtim on this quadrangle for exploration and comparison with the 2017 EHT data on \VirgoAstar.

\begin{deluxetable*}{lcccc}[ht]
\caption{Geometric Model Parameters}
\label{tab:params}
\tablehead{
\colhead{Parameter} & 
\colhead{Symbol} & 
\colhead{2017 Apr 5-6 Fiducial} & 
\colhead{2017 Apr 10-11 Fiducial} &
\colhead{Exploration Range\tablenotemark{e}}
}
\startdata
Annulus radius & $R$ & $20 \uas$ & $20 \uas$ & - \\
Annulus width  & $w$ & $5 \uas$ & $5 \uas$ & - \\
Stokes I Position Angle\tablenotemark{a} & $\Phi_{\rm I}$ & $0^\circ$ & $0^\circ$ & - \\
Polarization fraction & $m$ & 25\% & 20\% & - \\
Polarization Position Angle\tablenotemark{a} & $\Phi_{\rm pol}$ & $340^\circ$ & $330^\circ$ & - \\
Mean EVPA pitch angle\tablenotemark{b} & $\theta_0$ & $30^\circ$ & $30^\circ$ & - \\
Variable EVPA pitch angle\tablenotemark{b} & $\theta_1$ & $10^\circ$ & $10^\circ$ & $0$ to $20^{\circ}$ \\
Number of cycles\tablenotemark{c} & $n_{\rm cyc}$ & 0.183 & 0.183 & 0.183 to 270.29 \\
Axion phase offset\tablenotemark{d} & $\delta$ & $0^\circ$ & $0^\circ$ & $-90^{\circ}$ to $90^{\circ}$ \\
\enddata
\tablenotetext{a}{Starting from south, measured east of north.}
\tablenotetext{b}{Measured clockwise relative to radial direction.}
\tablenotetext{c}{Number of full periods in $24\,{\rm h}$, proportional to $\omega$, the axion angular frequency.}
\tablenotetext{d}{Defined such that $\theta=0^\circ$ at 0 UT on 2017 Apr 5 or on 2017 Apr 10, depending on which dataset we are fitting. 
}
\tablenotemark{e}{The ranges of parameters used for \autoref{fig:a0}, \autoref{fig:delta} and \autoref{fig:ncycles}.}
\end{deluxetable*}

\subsection{Sensitivity of CCTPs to Axion Signal}
\label{sec:Exploration}

A fiducial set of parameters, listed in \autoref{tab:params}, is chosen to qualitatively match the general morphology of the polarized images on 2017 April 11. \autoref{fig:Fiducial} shows the CCTP calculated from the fiducial parameters plotted against read CCTP of 2017 Apr 11. Specifically, we choose the position angle of the maximum brightness in Stokes $\mathcal{I}$ map to lie in the south, the position of the maximum polarized flux to lie in the southwest, and fixed polarization fraction of $m=25\%$.  The fiducial period is selected so that $\alpha \equiv r_{\rm g}/\lambda_c =0.4$, consistent with that shown in Figure 1 of \citet{Chen2020}, which corresponds to a period of roughly $131\,{\rm h}$, or $n_{\rm cyc}=0.183$ cycles in a $24\,{\rm h}$ period.  The CCTPs produced from this model match those observed on 2017 Apr 11 quantitatively.

\autoref{fig:a0} shows the impact of increasing the variable component of the EVPA pitch angle, $\theta_1$, associated with the axion cloud density.  For reference the 2017 Apr 11 high band CCTP measurements are shown in light gray, which indicate that CCTP-based constraints on $\theta_1$ of the order of $10^\circ$ should be possible.
In contrast, variations in $\theta_0$ (not shown) have no impact on the CCTPs; changing $\theta_0$ induces a total rotation in the EVPA across the image to which the closure traces are invariant.

Due to the long fiducial period, modifications of the axion phase offset $\delta$ can have a large impact on the sensitivity of CCTPs to a putative axion signal.  Shown in \autoref{fig:delta} are two models at very different locations in their respective oscillations.  The very different sensitivity to $\theta_1$ is easy to understand given the form of $\theta$ given in \autoref{eq:theta}.  When $\delta=0^\circ$, for small times $\theta$ evolves only slowly.  However, when $\delta=\pm90^\circ$, the evolution is maximized. For small times $\theta$ depends linearly on time, and thus the variation in the CCTPs is a combination of both $\theta_1$ and the temporal evolution throughout a single night.  Unsurprisingly, this behavior is strongly dependent on $n_{\rm cyc}$ as shown in \autoref{fig:ncycles}. Non-zero $\theta_1$ would induce peaks as substructures on the overall CCTP curves, with the number of peaks proportional to $n_{\rm cyc}$. In addition, for shorter periods, higher $n_{\rm cyc}$, the magnitude of the variation in the CCTPs driven by $\theta_1$ is generally increased, maximizing when $n_{\rm cyc}\gtrsim1$.  

The dependence on $\delta$ and $n_{\rm cyc}$ motivates a number of practical concerns for constraining the existence of axion clouds in EHT targets.  For \VirgoAstar, the large SMBH mass requires a large period, and thus small $n_{\rm cyc}$ --- constraints will benefit from coherently combining CCTP measurements across many nights.  This is further justified by the long intrinsic timescale for astrophysical changes in the emission region. 

However, for \SgrAstar, the much lower mass requires a smaller axion mass to drive the superradiant instability, and thus shorter orbital period and larger $n_{\rm cyc}$.  Simply rescaling by the SMBH mass and keeping other physical parameters the same, our fiducial \VirgoAstar model produce $n_{\rm cyc}=270.29$ for \SgrAstar, comparing to $n_{\rm cyc}=0.183$ for \VirgoAstar. This is shown in \autoref{fig:ncycles}, from which we conclude that even a single observation night may strongly constrain the existence of an axion cloud in the Galactic center.

\section{Implications of \VirgoAstar CCTP Data}
\label{sec:MCMC}

\begin{figure*}
    \includegraphics[width=\columnwidth]{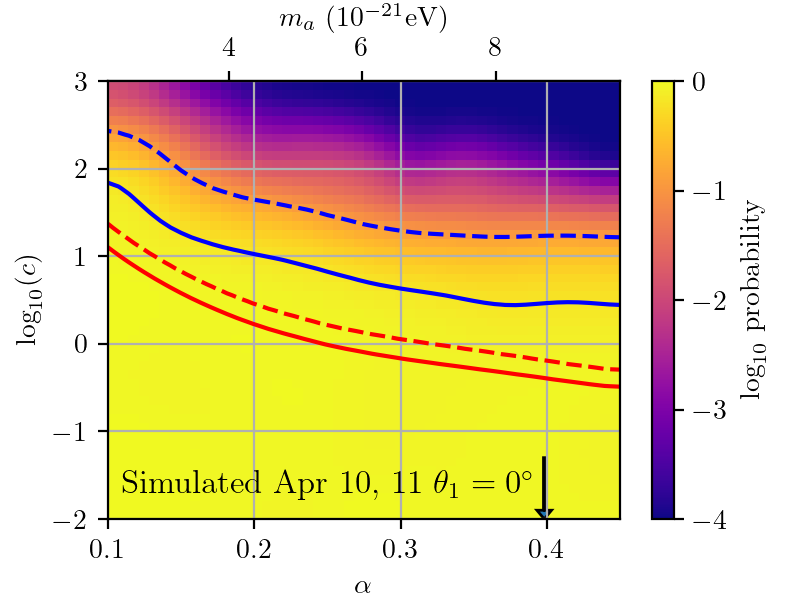}
    \includegraphics[width=\columnwidth]{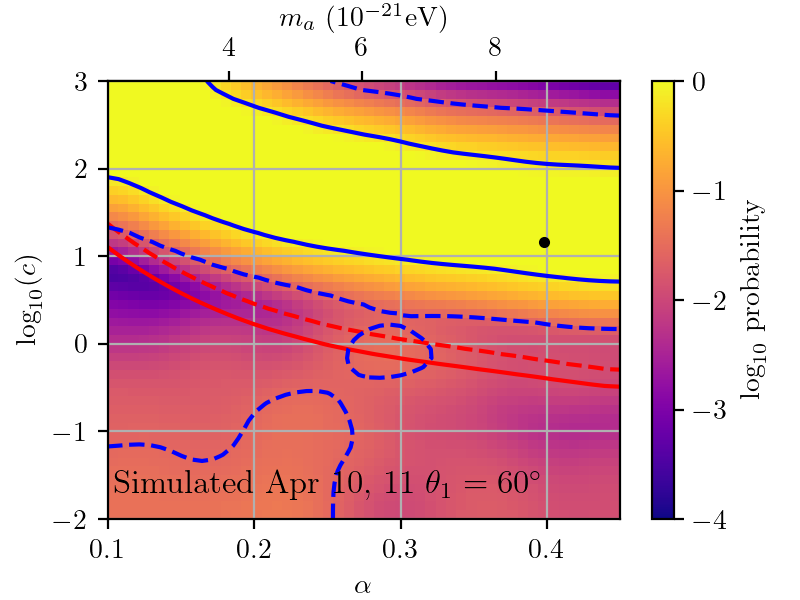}\\
    \includegraphics[width=\columnwidth]{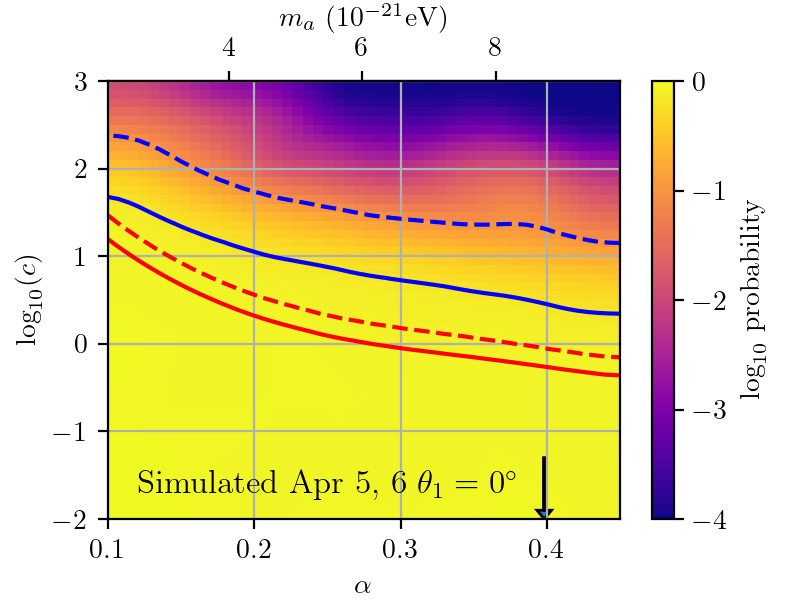}
    \includegraphics[width=\columnwidth]{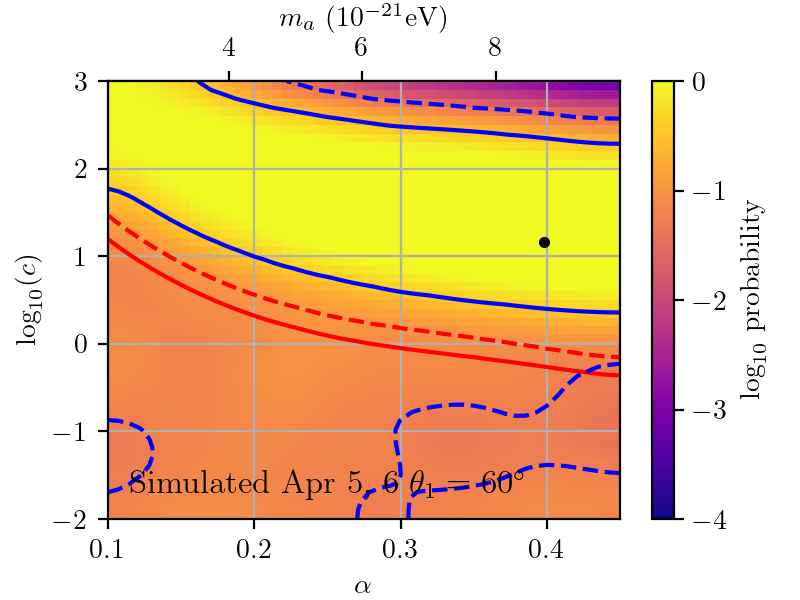}
    \caption{Posterior distribution plots for simulated data. The difference in date and predetermined $\theta_1$ is labeled at the bottom left. Constraints from \citet{Chen2022} are overplotted as red lines, while the blue lines are our estimates of excluded regions of axion parameters. Dashed lines are 3$\sigma$ constraints and solid lines are 2$\sigma$ constraints. The negative numbers on the colorbar imply a logscale of the probability in the corresponding posterior plot. The actual parameters used to generate those simulated cases are labeled as black dots on the posterior distribution plots or indicated by black arrows if they fall way below the lower limit.}
    \label{fig:posterior sim}
\end{figure*}

\begin{figure}
    \includegraphics[width=0.5\textwidth]{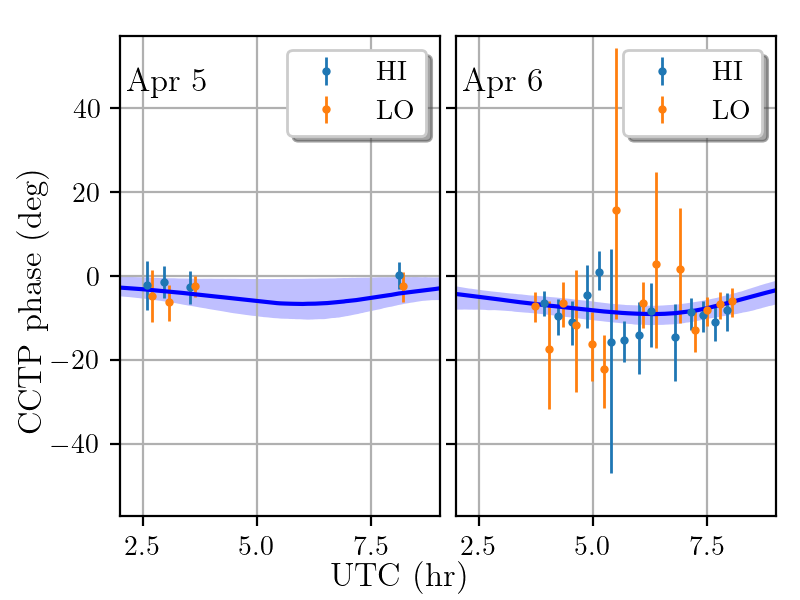}
    \includegraphics[width=0.5\textwidth]{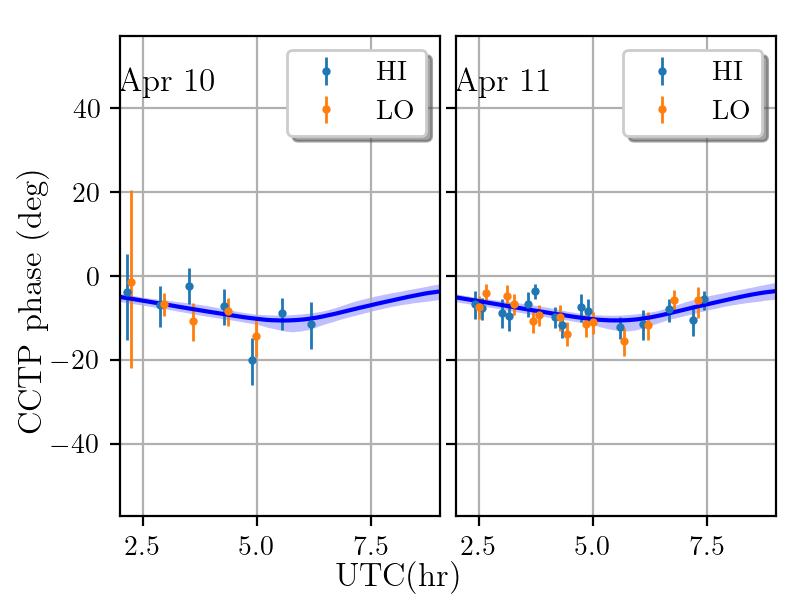}
    \caption{2$\sigma$ band plots over the real data, depicted in light blue, estimated from the MCMC fitting procedures on all 4 days. The central blue lines indicate the average fitting result.}
    \label{fig:comparison real}
\end{figure}

Armed with an expectation that CCTPs may be a sensitive discriminator among axion models in \VirgoAstar and \SgrAstar, we performed a number of fits to simulated and real data sets.  We begin with a summary of the fitting procedure, followed by demonstration on simulated data sets, and ending with an application to the 2017 EHT \VirgoAstar data.

\subsection{Fit Procedure}
In all cases, we make use of the Monte Carlo Markov Chain sampler \emcee to explore the posterior space of the model parameters \citep{emcee}. 
To fit the geometric model with the CCTPs of the entire 4 days of observations, both real and simulated data are grouped into two sets, each contains two consecutive days of observations, i.e Apr 5, 6 and Apr 10, 11, allowing the variability of \VirgoAstar in a timescale of a week. In addition, a fit directly to Apr 5, 6, 10, 11 is also performed. Beginning with the fiducial model parameters in \autoref{tab:params}, we calculate the corresponding CCTP values and then the log-probability by comparing the predicted and observed CCTP values, along with the corresponding errors, and summing over all datapoints, and update them within 200000 samples for a total of 40 walkers in parallel, i.e., a total of 8 million samples. Triangle plots are produced from chains \citep{corner}, and plots for the joint posterior distributions with axes converted to $\log(c_{a\gamma})$ and $\alpha$ in alignment with \citet{Chen2022} are generated as the axion limits derived from CCTP analyses. In particular, we transform the $\theta_1$-$n_{\rm cyc}$ plot into the $\log(c_{a\gamma})$-$m_a$ plot with the conversion factors discussed in the appendix. 

\newpage

\subsection{Simulated Data Sets}
In order to explore the impact of large $\theta_1$ and the significant errorbars that came with the observation data, a total of 4 simulated data sets were created based on the actual mass of \VirgoAstar and two choices of predetermined $\theta_1$. Posterior distribution plots for the simulated data sets are shown in \autoref{fig:posterior sim}. 

The null case ($\theta_1=0$) shows a clear excluded region of axion parameters in the upper right of the posterior plot, resembling the feature in Figure 4 of \citet{Chen2022}, although our boundary is weaker. However, the case with non-trivial $\theta_1=60^\circ$ indicates a significant clump of high probabilities located roughly between $\log(c_{a\gamma})=1$ and $\log(c_{a\gamma})=2$, where the actual parameters fall in. The "truth" is indeed narrowly constrained. Therefore, we believe the fitting works as expected since we clearly see anticipated behaviors in \autoref{fig:posterior sim}, which recovers reasonable values of \VirgoAstar mass and $\theta_1$, and uncertainties. Overall, considering the fact that simulated data sets were constructing based on the method of control variable, patterns deviating from the null case posterior plot can indicate a possible detection of light axions that can trigger EVPA variations with magnitude above the level which can be explained by the errorbars. 

\subsection{2017 EHT \VirgoAstar Campaign}
\begin{figure}
    \includegraphics[width=\columnwidth]{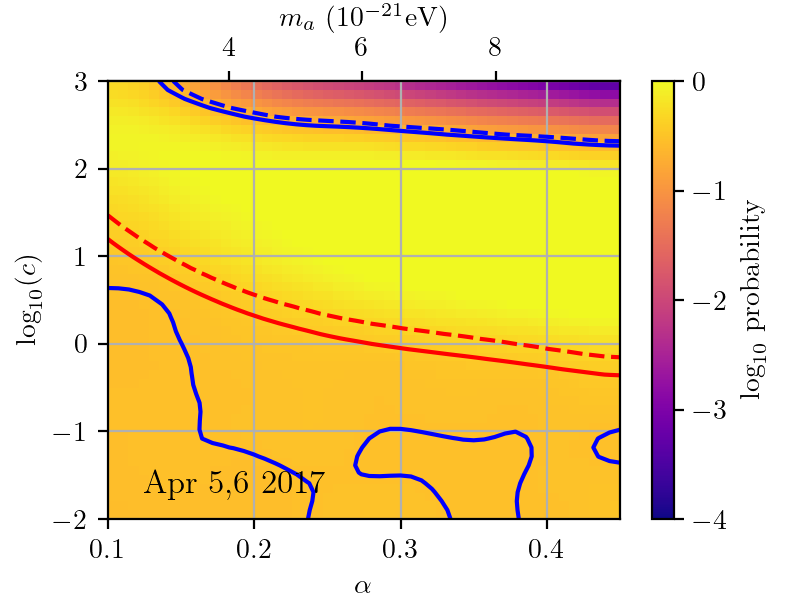}
    \includegraphics[width=\columnwidth]{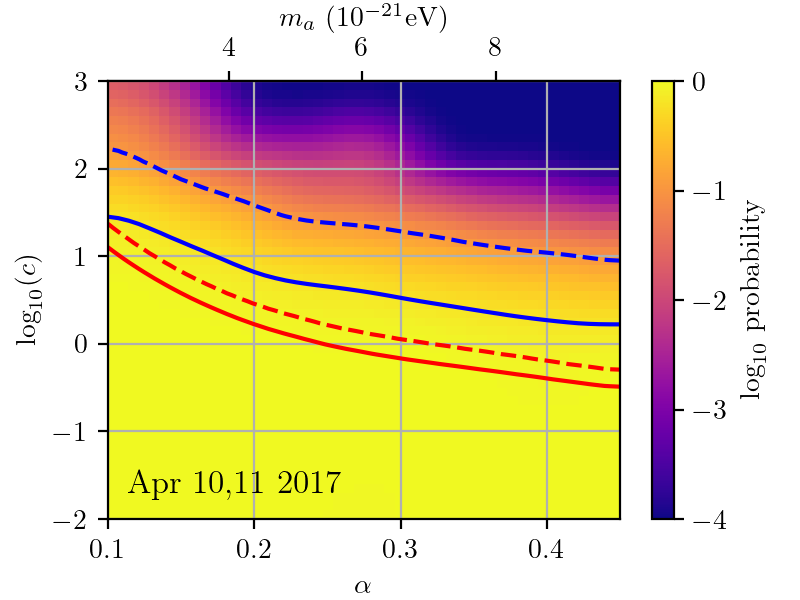}
    \includegraphics[width=\columnwidth]{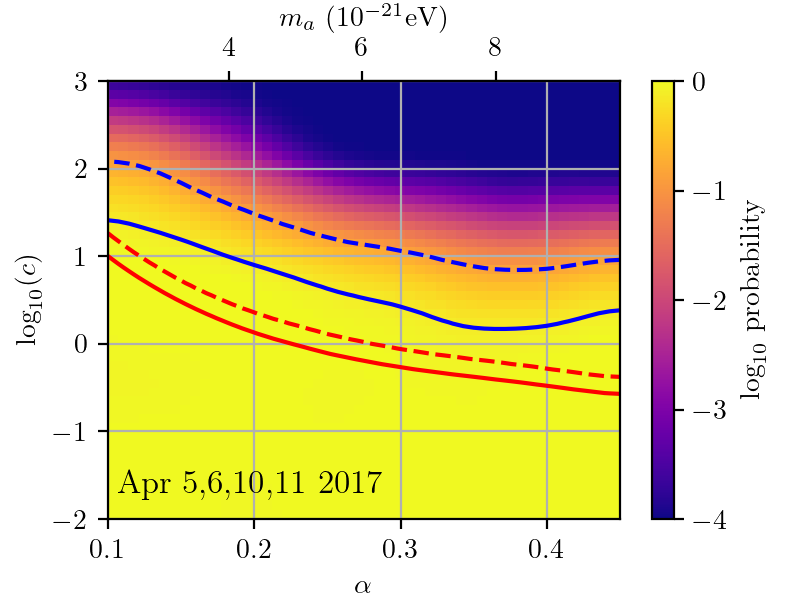}    
    \caption{Posterior distribution plots for real data. Constraints from \citet{Chen2022} are overplotted as red lines, while the blue lines are our estimates of excluded regions of axion parameters. Dashed lines are 3$\sigma$ constraints and solid lines are 2$\sigma$ constraints. The negative numbers on the colorbar imply a logscale of the probability in the corresponding posterior plot. }
    \label{fig:posterior real}
\end{figure}

As stated above, fits to EHT data were performed on the 2017 Apr 5, 6 and Apr 10, 11 data sets separately. In addition, a fit to the entire data set (2017 Apr 5, 6, 10, 11) was also executed. The resulting 2$\sigma$ band of fits is overplotted on the EHT data in \autoref{fig:comparison real}. 
Posterior distribution plots, including 2$\sigma$ and 3$\sigma$ regions, are shown in \autoref{fig:posterior real}. The posterior plot for Apr 10, 11 clearly matches a non-detection of light axions at even 2$\sigma$ and is consistent with the null case posterior plot in \autoref{fig:posterior sim}. Although it may seem that on Apr 5, 6, there is a phantom of detection at 2$\sigma$ level, there is no evidence for any detection at 3$\sigma$. This interpretation, is confirmed by the fit to the entire data set.
That is, we see no evidence for an axion cloud from the 2017 \VirgoAstar EHT campaign.

\subsection{Discussion}

In summary, fits for simulated data sets properly recover the undelying predetermined parameters, indicating the feasibility of using CCTP measurements to probe for light axions. For EHT 2017 data, at 3$\sigma$ level, we conclude that there is no detection of light axions and therefore we arrive at an upper limit on the putative axion-photon coupling. 

It is essential to carefully define what a ``detection'' and ``non-detection'' of axions means in this context given the substantial astrophysical uncertainties. For a non-detection it suffices to conclude that within the current accuracy that the null hypothesis, no axion cloud, is consistent with the observation. Consequently, we exclude the axion parameter spaces above 
an observationally-dependent bound, above which any light axion is expected to result in CCTPs that are inconsistent with the EHT data.  Conversely, the EVPA variations due to lower density axion clouds are consistent within the formal uncertainties. 

We note that to claim a ``detection'' of an axion cloud, the elimination of the null hypothesis is necessary but insufficient. In addition, all potential astrophysical origins for the EVPA oscillations (e.g., evolving magnetic field structures) must be excluded, something we have not explored here. Therefore, at present we can only produce upper-limits on the light axion cloud density associated with non-detections, and therefore the coupling (subject to assumptions about the history of the axion cloud).

\begin{figure}
    \includegraphics[width=\columnwidth]{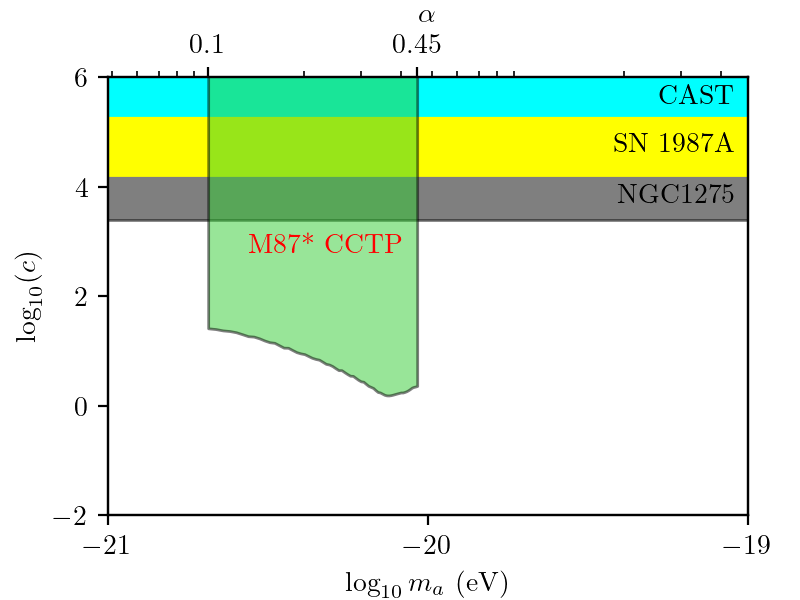}
    \caption{The 90\% limit on the axion-photon coupling(lime-green), set by the fit to the entire dataset (Apr 5, 6, 10, 11), estimated from the polarimetric data in EHT \VirgoAstar 2017 campaign via CCTP analysis. In comparison, by assuming $f_a = 10^{15} {\rm GeV}$, bounds from CAST\citep{CAST_limit}, SN1987A\citep{SN1987A_limit} and NGC1275\citep{NGC1275_limit} are also shown.}
    \label{fig:mylim}
\end{figure}

\autoref{fig:mylim} summarizes our constraint to light axion parameters derived via the CCTP analysis on the \VirgoAstar 2017 linear polarization data. The bound on the dimensionless coupling constant becomes weaker for smaller axion masses, primarily due to a smaller value of the radial wave function of the axion cloud and a smaller single-day axion field variation resulting from the longer oscillation period \citep{Chen2022}. Despite presenting a modestly weaker limit than the preceding one, our result highlights the feasibility of CCTP analyses for constraining the existence of axion clouds. More importantly, the image-based analysis of \citet{Chen2022} is inevitably subject to significant underlying systematic uncertainties originating from the estimation of station gains and D-terms, to which the CCTP-based analysis is insensitive. Additionally, the use of CCTP analyses on the polarization data of other SMBHs, for which full polarimetric imaging may be infeasible or impractical, holds potential for cross-validation of the existence of light axions; more is discussed in \autoref{sec:Conclusion}.

\section{Conclusion}
\label{sec:Conclusion}

Photons propagating in a light axion background around black holes would experience a variable, spatially-dependent birefringence effect, caused by axion-photon coupling, which may be detectable via polarimetric studies of the horizon-scale emission. Image-based studies on constraining the parameters of light axion have been done by modeling the temporal EVPA variations around the supermassive black hole \VirgoAstar. 

Unlike previous efforts, CCTPs provide a novel non-imaging method for constraining the existence of light axions.  The direct analysis of closure quantities, i.e., closure traces and the CCTPs generated from them, provide two significant advantages.  First, they are intrinsically immune to station-based polarimetric systematics, including atmospheric phase delays and absorption, and polarimetric leakage.  Second, they are a non-imaging quantity, and therefore do not require image reconstructions and their attendant assumptions.

Application of CCTP-fitting to EHT observation of \VirgoAstar finds no detection of axions, resulting in an upper limit on the coupling constant of $\log_{10}(c_{a\gamma}) \lesssim 1$ for axions with mass $2.2\times 10^{-21} {\rm eV}<m_a<9.9\times 10^{-21} {\rm eV}$. 
This conclusion is, however, based on the assumption that a saturated axion cloud exists around \VirgoAstar, generated by the superradiant instability and limited only by axion-axion annihilation (bosenova).  That is, we ignore the complicated astrophysical history of this target.

A significant current uncertainty in the \VirgoAstar analysis is the unknown phase of the axion cloud ($\delta$ in \autoref{eq:theta}).  Future observations of \VirgoAstar extending over multiple weeks may retire this uncertainty, though potentially at the expense of introducing additional astrophysical variability. 

Applying this technique to \SgrAstar would probe a very different range of axion masses, and therefore temporal scales.  The expected CCTP signal is qualitatively different for \SgrAstar than \VirgoAstar, exhibiting rapid, periodic fluctuations throughout a single night that may be excluded by existing and future EHT observations; tests based on \SgrAstar are forthcoming. Because the generation of CCTPs does not require full imaging, our method may be applied to marginally resolved and possibly unresolved targets, including many AGN targets beyond \VirgoAstar and \SgrAstar, and thereby expanding the range of axion masses probed.

Already, following the 2017 EHT campaign, additional sites have been added to the array, increasing the number of available quadrangles significantly. The improved baseline coverage, spatial and temporal resolution of the next-generation EHT \citep{ngEHT} will enable more detailed geometric modeling and more precise EVPA measurements, resulting in stronger constraints on light axions.  The magnitude of the improvement can be roughly estimated from the increased volume of observations ($4\,{\rm d}$ to $16\,{\rm d}$), increased bandwidth ($4\,{\rm GHz}$ to $64\,{\rm GHz}$), increased number of quadrangles (20 stations will produce 170 independent quadrangles, in comparison to the single quadrangle considered here), and by analyzing both the amplitude and phase of the CCTPs on these quadrangles.  Assuming that each quadrangle produces CCTPs with similar statistical power, the limit on the axion-photon coupling constant, $\log_{10} (c_{a\gamma})$ could be improved by two orders of magnitude.  Verifying this with simulated data sets is a natural future project.

\begin{acknowledgments}
We thank Asimina Arvanitaki and Yifan Chen for their valuable discussions. This work was supported in part by Perimeter Institute for Theoretical Physics. Research at Perimeter Institute is supported by the Government of Canada through the Department of Innovation, Science and Economic Development Canada and by the Province of Ontario through the Ministry of Economic Development, Job Creation and Trade. A.E.B. thanks the Delaney Family for their generous financial support via the Delaney Family John A. Wheeler Chair at Perimeter Institute. A.E.B. receives additional financial support from the Natural Sciences and Engineering Research Council of Canada through a Discovery Grant.
\end{acknowledgments}

\software{\ehtim, \emcee}

\begin{appendix}

\section{Appendix} \label{app:detrending}
While natural to characterize the observational impact of the axion cloud on the evolving EVPA in terms of an amplitude ($\theta_1$) and number of cycles ($n_{\rm cyc}$), these quantites are directly related to the underlying axion properties.  Here we collect the various conversion relations between the two sets of quantities.

The oscillation frequency of the axion field, $\omega_a$, is proportional to the axion mass, $m_a$ \citep{Plascencia2018}.  Incorporating fundamental constants, the number of cycles in a $24\,{\rm h}$ period is then related to $m_a$ via,
\begin{equation}
    n_{\rm cyc} = \frac{\omega_a}{2\pi} 24\,{\rm h} = \frac{m_a c^2}{h} 24\,{\rm h}.
\end{equation}
Alternatively, we can express $m_a$ in terms of the axion fine structure constant, $\alpha$, by
\begin{equation}
    \alpha = \frac{r_g}{\lambda_c} = \frac{r_g m_a c}{\hbar},
\end{equation}
and therefore,
\begin{equation}
    n_{\rm cyc} = \frac{\alpha c}{2\pi r_g} 24\,{\rm h} \approx 0.458\alpha,
\end{equation}
where $r_g$ for \VirgoAstar was inserted.

The conversion between $\theta_1$ and $c_{a\gamma}$ employs the approximation that the amplitude of the EVPA variation is only weakly dependent in $\phi$, an assumption that is supported by Figure~2 of \citet{Chen2022a}.  We denote the azimuthally averaged amplitude from \citet{Chen2022a} by $\langle\mathcal{A}(\phi)\rangle_{\phi}$.  It is convenient to normalize this by the axion coupling constant and maximum axion field strength,
\begin{equation}
    \mathcal{A}'(\phi) \equiv \frac{\mathcal{A}(\phi)}{g_{a\gamma}a_{max}}.
\end{equation}
At saturation, the axion field strength is related to the dimensionless axion coupling constant by $c_{a\gamma} \equiv 2\pi g_{a\gamma} a_{max}$.  Therefore, $\theta_1$ and $c_{a\gamma}$ are related by
\begin{equation}
    \theta_1 = \langle\mathcal{A}(\phi)\rangle_{\phi} = \frac{\langle\mathcal{A}'(\phi)\rangle_{\phi}}{2\pi} c_{a\gamma}.
    \label{theta(c)}
\end{equation}
The specific values of $\mathcal{A}'(\phi)\approx0.5$ though depend on $\alpha$, thus \autoref{theta(c)} are evaluated for a list of evenly spaced $\alpha$ from 0.1 to 0.45 assuming an accretion flow thickness of $H=0.3$.

\end{appendix}

\bibliographystyle{yahapj}
\bibliography{references,EHTCPapers}

\end{document}